\documentclass[12pt]{amsart}
\usepackage{ljm-auth}
\usepackage{graphicx}

\author{N.A.\,Inogamov$^{1,2}$, V.V.\,Zhakhovsky$^{2,1}$   
}
\crauthor{INOGAMOV and ZHAKHOVSKY} 

\tit{Simulations of short pulse laser-matter interaction}
\shorttit{Short pulse laser ablation} 

\begin{document}

\maketit

\address{$^1$Landau Institute for Theoretical Physics, Russian Academy of Sciences, Chernogolovka 142432, Russia}
\address{$^2$Dukhov Research Institute of Automatics, Rosatom, Moscow 127055, Russia}


\Received{date}

\abstract{Studies of ultra-fast laser-matter interaction are important for many applications.
 Such interaction triggers extreme physical processes
  which are localized in $\sim$10 nanometers$\div$micron spatial scales
   and developing within picosecond--nanosecond time range.
 Thus the experimental observations are difficult and methods of applied mathematics are necessary
   to understand these processes.
 Here we describe our simulation approaches and present solutions for a laser problem significant for applications.
 Namely, the processes of melting, a liquid jet formation, and its rupture are considered.
 Motion with the jet is caused by a short $(\sim 0.1\div 1$ps) pulse illuminating a small spot
  on a surface of a thin $(\sim 10\div 100$nm) film deposited onto substrate.
  }

\notes{0}{
 \thank{This work was supported by Russian Science Foundation (No. 14-19-01599)} }

\section{Introduction}
\label{sec1:intro} Laser technologies are widely used in modern industry.
 Our studies of blistering of a thin film, appearance of a jet, and its break off are related to
   the LIFT (laser induced forward transfer) and nanoplasmonic technologies
     \cite{Korte:2004,Unger:2012,Zywietz:2014,Kuchmizhak:2016,Ivanov:2013}
       and to technologies of creation of superfine structures and functional surfaces \cite{Nakata:2009,Nakata:2013}.
 Nakata and coworkers have developed an interferometric technique combining few laser beams to produce 2D surface arrays
   composed from the solitary microbumps \cite{Nakata:2009,Nakata:2013}.
 These arrays are used for the SERS (surface enhanced Raman scattering), for enhancement of photoluminescence, and in nanophotonic devices
  \cite{Zywietz:2014,Kuchmizhak:2016,Nakata:2009,Nakata:2013}.

 As in experiments \cite{Korte:2004,Unger:2012,Zywietz:2014,Kuchmizhak:2016,Ivanov:2013,Nakata:2009,Nakata:2013},
  we consider action of tightly focused laser beam onto a thin film deposited on a glass substrate.
 Minimal sizes of illumination spot are achieved at the diffraction limit.
 For optical light this means that the minimal size is $\sim 1$ micron.
 Detailed distribution of intensity over the spot depends on numerical aperture of focusing system.
 It is important that a pulse is short, it is shorter than duration $t_{eq} \sim 5$ ps of the electron-ion temperature equalization process
   in gold \cite{Inogamov:2015a} used as material for a film.
 We consider experiments with duration of a laser pulse $\tau_L \sim 0.1\div 1$ ps.
 Therefore absorption of a pulse is accompanied and continued by two-temperature relaxation
   which forces electron $T_e$ and ion $T_i$ temperatures together: $T_e\approx T_i;$
    during a pulse the electrons absorbing laser energy are much hotter: $T_e\gg T_i.$

 Another and even more significant duration is an acoustic time scale $t_s=d_f/c_s,$
   where $d_f$ and $c_s$ are thickness of a film and speed of sound in a film.
 We consider thin films, their thickness $d_f$ is thinner than thickness $d_T$ of a heat affected zone; $d_T=100\div 200$ nm for gold.
 In this case an electron heat conduction (ehc) equalizes electron temperature $T_e(r,z,t)$ spatially along thickness of a film,
   i.e. along the direction $z$ normal to the initially plane film, here $z$ is the axis of a laser beam, $r$ is cylindrical radius.
 In gold the $z$-equalization of $T_e(r,z,t)\to T_e(r,t)$ is fast, it is even faster than the electron-ion temperature equalization: $t_{ehc-z}<t_{eq}.$
 Conductive radial spreading of thermal energy is much slower process relative to the hierarchy of durations:
  $\tau_L\sim 0.1-1$ ps, $t_{ehc-z}\sim 1$ ps, $t_{eq}\sim 5$ ps, $t_s\sim 10\div 30$ ps for $d_f=30\div 100$ nm.
 Duration of the radial cooling is $t_{frz}\sim R_L^2/\chi,$ where $\chi\sim 1$ cm$\!^2/$s is thermal diffusivity,
   $t_{frz}\sim 10$ ns for laser beam diameters $2R_L\sim 0.5\div 2$ microns; we neglect heat conduction of a dielectric substrate.
 The time scale $t_{frz}$ defines a rate of freezing of a focal spot molten by laser action.
 In the paper we restrict ourself to the practically important problem of the nanobumping of a thin film $d_f=30\div 100$ nm
   under action of a diffraction limited laser irradiation $(R_L\sim 1$ micron) by an ultrashort pulse $\tau_L\sim 0.1\div 1$ ps.

\section{Separation of a film from substrate}
\label{sec2:kick-off.SEPARATION}
 Separation of a film from substrate is a result of the thermomechanical kick-off.
 Fast absorption of laser energy $F_{abs}$ sharply increases pressure
  \begin{equation} \label{eq:1-Pf}
 p_f\approx\Gamma\, F_{abs}/d_f = 0.3\,F_{mJ}/d_{100} {\rm [GPa]}
  \end{equation}
  in a film,
   here $\Gamma\approx 3$ is Gruneisen parameter for gold, $F_{mJ}=F_{abs}/(1$ mJ/cm$\!^2),$ $d_{100}=d_f/(100$ nm).
 In our case when $\tau_L\ll t_s$ the supersonic injection of energy into a film takes place.  
 This means that gold is heated {\it before} it has time to expand to the equilibrium volume $V_{bin}(T)$ corresponding to the increased temperature $T.$
 Pressure in the volume $V_{bin}(T)$ equals to the saturation vapor pressure $p_{sat}(T).$
 In our range of absorbed fluences $F_{abs}=10\div 100$ mJ/cm$\!^2$ the pressure $p_f\approx 0.3\,F_{mJ}/d_{100}$ [GPa] is much higher than $p_{sat}(T).$

 There is a contact surface between a film and a substrate.
 The pressure $p_{cb}(r,t)$ at the contact boundary (cb) rises up almost simultaneously with the laser absorption in a skin layer $\delta_{sk}<d_f,$
  because $t_{ehc-z} \ll t_s;$ $\delta_{sk}\sim 10\div 20$ nm for optical lasers.
 There are some fine features in the rise of total pressure $p=p_e+p_i$ connected with energy transfer from electrons to ions \cite{Inogamov:2015a}.
 They follow from the fact that the ion Gruneisen parameter equals $\Gamma_i\approx 3$ while the electron Gruneisen parameter is less than 1
  (electron subsystem is softer).
 Therefore addition of the internal energy per unit of volume delivered into electron subsystem causes significantly lower increase
  of the total pressure $p=p_e+p_i$
   than increase of pressure $p=p_e+p_i$ due to addition of the same amount of energy into the equilibrium system $T_e=T_i=T.$
 This is true because the electron heat capacity $\sim (T_e/T_F)\, k_B$ is much less than the ion heat capacity $\approx 3 k_B$
   in our range of the equilibrium temperatures $T=T_e=T_i.$
 In this energy range the equilibrium temperature $T=T_e=T_i\sim 1.5\div 3$ kK is small relative to the Fermi temperature $T_F.$

 We have
  \begin{equation} \label{eq:2-Pcb}
 p_{cb}\approx p_f\, Z_g/(Z_A+Z_g)\approx 0.13\, p_f = 0.04\,F_{mJ}/d_{100} {\rm [GPa]}
  \end{equation}
 where $Z_g$ and $Z_A$ are acoustic impedances $Z=\rho\, c_s$
  of glass substrate and gold (Au), respectively; $Z_g/Z_A\approx 0.15$.
 Thus $p_{cb}$ is significantly less than $p_f.$
 Nevertheless, pressure $p_{cb}$ is much higher than pressure of saturation $p_{sat}(T)$ in our range of the temperatures $T=T_e=T_i=1.5\div 3$ kK.
 Therefore evaporation cannot influence dynamics of film separation.

 Gold is weakly coupled to the glass substrate.
 In this case a film separates from substrate as the rarefaction wave propagating from the film/vacuum boundary achieves the contact.
 The propagation time for the rarefaction wave is $t_s.$                   
 Let us neglect variation of the contact pressure $p_{cb}$ during the two-temperature case, because $t_{eq}<t_s;$
    $t_{eq}\sim 5$ ps, $t_s\sim 10\div 30$ ps for $d_f=30\div 100$ nm.
 Then balancing momentum $p_{cb} \, t_s$ created by reaction of a substrate to expansion of gold
   and the momentum $\rho_A\, d_f \, u_f$ of a film after separation
    we find final velocity of a film after separation from substrate:
 \begin{equation} \label{eq3:u=coef*Fabs}
 u_f = p_{cb}/\rho_A\, c_s = 0.6 \, F_{mJ}/d_{100} {\rm [m/s],}
 \end{equation}
  where $\rho_A$ is density of gold.
 Namely the reactive momentum $p_{cb} \, t_s$ is responsible for the laser kick-off of a film.

\section{Capillary forces and reverse of inflation of cupola to deflation}
\label{sec3:Capillary}
 According to relation (\ref{eq3:u=coef*Fabs}) the radial velocity distribution $u_f(r)$ repeats the fluence distribution $F_{abs}(r)$
  for a film of homogeneous initial thickness $d_f\equiv$const. 
 Usually the distribution $F_{abs}(r)$ is a smooth Gaussian type function with a maximum in the center of a focal spot.
 If a film is molten and $R_L\gg d_f$ then the lateral velocities created by the kick-off are small relative to the velocities in normal direction
   and the separated film moves as $z(r,t)\approx u_f(r) \, t$
     forming 
      a bulging, moving shell similar to cupola; here time $t$ is reckoned from the instant of arrival of an ultrashort laser pulse;
        we say "instant" of arrival because duration of a pulse $\tau_L\sim 0.1\div 1$ ps is much shorter than the nanosecond temporal scale $t_{ns}$
         when the bulging of a film becomes appreciable: $z(r,t_{ns})\sim R_L.$

 The edges of the cupola remains mechanically and thermally coupled to the rest of a film around the illuminated spot.
 A film outside the spot keeps its initial plane geometry and weak adhesion to the substrate.
 Position of the edge of the cupola is defined by the value of weak adhesion (a threshold of separation)
   or/and the boundary between the solid and molten parts of a film
      because it is much more difficult to bend a solid film than a liquid one.
 There is significant bending of a film near the edge of cupola.

 The molten cupola inflates thanks to the kick-off velocities and inertia of its mass.
 In the cases interesting for us, the capillary deceleration of the inertial flight becomes significant
   when the curvature $1/r_{curv},$ $r_{curv}\sim R_L^2/2 z(0,t)$ of the cupola becomes moderately large $r_{curv}\sim R_L;$
     here function $z(r,t)$ gives the cupola.
 Capillary pressure $p_{cap} = 2\sigma/r_{curv}$ grows with a curvature; here $\sigma$ is coefficient of surface tension.
 Value $p_{cap}\sim 20$ bar for $\sigma=1000$ dyne/cm$\!^2$ and $r_{curv}\sim 1$ micron.
 It is higher than saturated vapor pressure $p_{sat}$ in conditions where structures with cupola or cupola plus jet are formed;
   boiling temperature of gold is $T_b=3078$ K, $p_{sat}(T_b)=1$ bar.
 At what stage of inflation the pressure $p_{cap}$ becomes significant?
 This depends on the value of the surface tension parameter $\xi_{cap} = v_\sigma/u_f(0),$
  where capillary velocity $v_\sigma=2\sqrt{\sigma/\rho_A\, d_f}$$=$$45/\sqrt{d_{100}}$ m/s (for $\sigma=1000$ dyne/cm)
    is defined by comparison of kinetic and surface energies: $\rho_A\,d_f\, v_\sigma^2/2=2\sigma;$
      $u_f(0)$ is the kick-off velocity in the center of a laser beam.
 It is clear that if the coefficient $\sigma$ is large then the capillarity becomes significant early.
 Let's consider the question about competition between inertia and surface tension.

 Kinetic energy $\rho_A\,d_f\, u_f(0)^2$ is finite while the surface energy increases infinitely with stretching of surface:
   $2\sigma\, d_f(0)/d_f(t),$ \{where $d_f(0)$ and $d_f(t)$ are initial and current thicknesses of a film;
    the factor $d_f(0)/d_f(t)$ follows from conservation of mass $dS(0)\, d_f(0)\, \rho(0) = dS(t)\, d_f(t)\, \rho(t);$
  where $dS$ is surface of a small part of a cupola shell, $\rho(0) \sim \rho(t)$\}.
 Therefore formally in any case the surface tension will stop inflation and will return all separated mass back onto substrate.

\begin{figure}[h]    
\includegraphics[width=1.0\textwidth]{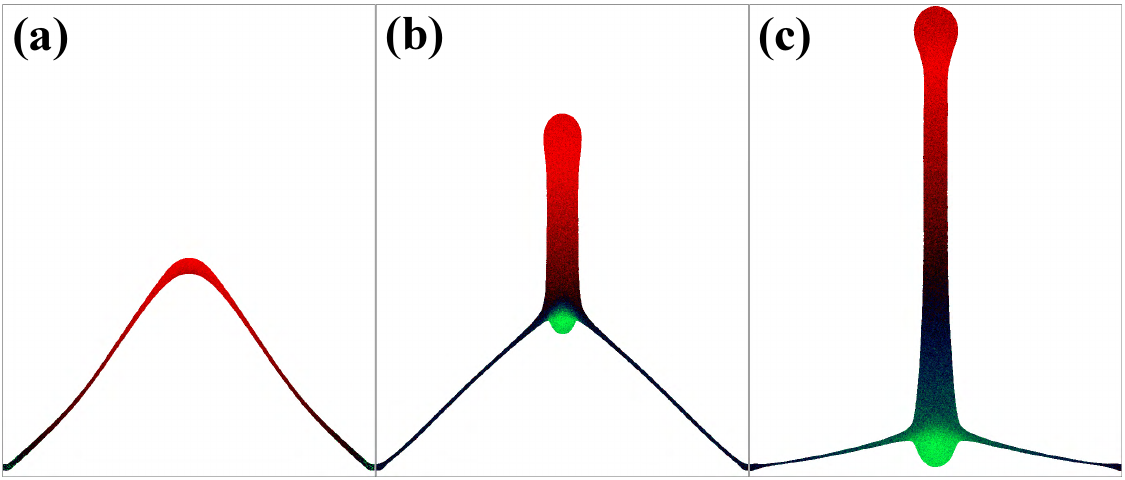}  
\caption{Kick-off inflation, after that capillary deceleration, deflation, and formation of the jet.
 (a) Formation of cupola after separation of a film from substrate,
 $\hat t=t/t_\sigma=0.37;$ $t_\sigma=R_L/v_\sigma=2.6$ ns.
 Run 203: $\xi_{frz}=0$ - we neglect freezing in this run;
   $\xi_{cap}=0.38,$ $R_L=300$ nm, $d_f=8.3$ nm,
     $\sigma=540$ dyne/cm for our EAM (embedded atom method) interatomic potential of gold \cite{Zhakhovsky:2016a},
       $v_\sigma=116$ m/s.
 Formation of cupola takes place under action of surface tension.
 (b) Appearance of a jet and a counter-jet after beginning of deflation of the cupola, $\hat t=0.997.$
 Red -- vertical velocity up, green -- down.
 (c) Strong elongation of the jet $\hat t=1.62.$
 The counter-jet remains small relative to the jet. Comp. with Fig. 3a from \cite{Nakata:2013}
   where also the counter-jet fills the gap between the tip of the cupola and the substrate.
     }
      \label{fig:1-203}
\end{figure}

 The stage when significant deflation and the stop of cupola will take place depends on $\xi_{cap}.$
 There are cases with early $\xi_{cap}\gg 1,$ middle $\xi_{cap}\sim 1,$ and late $\xi_{cap}\ll 1$ stopping:
   $z_{stop}\ll R_L,$ $z_{stop}\sim R_L,$ $z_{stop}\gg R_L.$
 There are three factors that intervene in the inertia/tension competition in real situation.
 They are (I) melting-separation against adhesion to substrate, (II) freezing of a molten circle of a film,
   and (III) rupture processes.

 (I) Adhesion shifts the kick-off velocity (\ref{eq3:u=coef*Fabs}) $u_f=p_{cb}/Z_A$ $\to$ $u_f\approx p_{cb}/Z_A - u_{adh}.$
 The last expression is valid for velocities above the threshold $u_f > u_{adh},$ where the threshold $u_{adh}$ increases proportionally to an adhesion strength.
 We consider the cases with weak adhesion, see example in Fig. \ref{fig:1-203}.
 This means that $u_{adh}\ll u_f|_m,$ where the subscript "m" relates to melting.
 The melting thresholds on absorbed energy $F_{abs}|_m$ and on kick-off velocity (\ref{eq:1-Pf}-\ref{eq3:u=coef*Fabs}) are
 $$
 F_{abs}|_m = [ 3 \,k_B\, n (T_m-T_{rt}) + Q_m ] d_f = 38\, d_{100} \, [{\rm mJ/cm}^2];\,  u_f|_m=23 \, [{\rm m/s}],
 $$
 where $Q_m,$ $T_m,$ $T_{rt}$ are heat of fusion and temperatures of melting and room conditions.
 The $F_{abs}|_m$ is proportional to thickness (till $d_f<d_T),$
   while $u_f|_m$ is quantity independent of energy $F_{abs}$ or thickness $d_f.$

 Adhesion may be made large by introducing an intermediate chromium layer between gold and glass
   as it was done in work \cite{Nakata:2009}.
 In the case of large adhesion the situations with cupola plus a jet \cite{Unger:2012,Nakata:2013,Inogamov:2015a}
  disappear and only situations with a hole in a film become possible at rather high fluences.
 Indeed, we have to adjust velocity $u_f(0)$ above the melting threshold $u_f|_m$ but value $u_f(0)$ should be less than the value $\sim v_\sigma$
    to create the cupola with a jet.
 Obviously strong adhesion $u_{adh}\gg u_f|_m$ opposes the conditions $u_f|_m < u_f <\sim v_\sigma$
   which limit rather narrow range.

 (II) The return back of a flying film under capillary action may be canceled by recrystallization of liquid gold.
 Namely freezing into a solid state allows appearance of the final cupolas \cite{Inogamov:2016a,Inogamov:2016b}.
 Solid mechanically behaves very differently relative to liquid.
 Solidification suppresses stretching of the cupola shell and jet.
 Rate of cooling is defined by thermal velocity $v_{frz}=\chi/R_L$ $=$ $100\, \chi_1/(R_L)_1$ m/s,
   where $\chi\sim 1$ cm$\!^2/$s is a heat diffusion coefficient for gold, $\chi_1=\chi/(1$ cm$\!^2/$s),
     $R_L|_1=R_L/(1$ micron).
 Thus a thermal parameter $\xi_{frz} = v_{frz}/u_f(0)$ is added to the capillary number $\xi_{cap}$
  due to importance of freezing.

 These parameters are $\xi_{cap}\sim 1,$ $\xi_{frz}\ll 1$ for the situation shown in Fig. \ref{fig:1-203}
  (and adhesion is weak).
 This is the situation where or $R_L\sim 1$ micron and $\chi$ is small
  relative to the usual values $\sim 1$ cm$\!^2/$s,
   or $R_L$ is large (a few or many microns) and $\chi$ is usual.
 Then a hole in a film appears.
 A hole is surrounded by a rim made from frozen remnants sometimes called nanocrown.
 A molten cupola shell of large radius $R_L$ decays into large droplet or into many smaller droplets
  before it will be stopped by surface tension.
 Decay of a large shell into small droplets is governed by amplification of the surface density [g/cm$\!^2]$ inhomogeneities.

 (III) Rupture processes are complicated.
 There are cases with volume, surface, or rod break-off.
 In the second and the third cases these processes are caused by strong stretching down to atomic scale or by inhomogeneities and instabilities.
 The first case (nucleation in volume) has been considered in many papers,
     see, e.g., \cite{Inogamov:2014c,Wu:2016a,Mayer:2016}.  
 Rupture of a cupola shell has been studied in paper \cite{Inogamov:2016a} (this is the case \#2).
 Below we present new results concerning rupture of a jet (the case \#3).

 \section{Development of Software}
 \label{sec4:soft}

 Special software was created to study the problem of laser formation of surface structures.
 The kick-off part (see Section \ref{sec2:kick-off.SEPARATION}) of the problem is simulated using two-temperature hydrodynamics code \cite{Inogamov:2015a}.
 This code transfers a distribution of fluence $F_{abs}(r)$ over an illuminated spot into the distributions of temperature $T(r)$
  and normal component of velocity $u_f(r)$ of a film at the instants of separation of a film from a glass substrate.
 At the stage of separation the two-temperature stage is finished, thus the one-temperature equations may be used.
 The distributions of $T(r)$ and $u_f(r)$ are employed in the molecular dynamics (MD) code as the initial data.
 Thus the MD code picks up the evolution of a film on substrate after separation of a film in the irradiated circle
  and follows further the inflation, deflation, and freezing parts of the problem.

\begin{figure}[h]    
\includegraphics[width=1.0\textwidth]{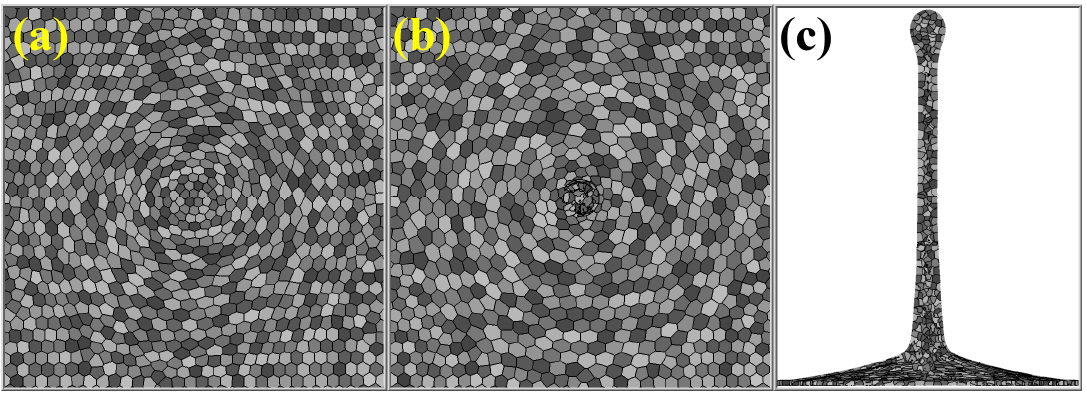}
\caption{ Problem of microstructure formation is solved using two-temperature hydrodynamics Lagrangian one-dimensional code
 which is combined with our parallel molecular dynamics (MD) code using Voronoi dynamic domain decomposition $(VD^3)$ method.
 Therefore we use multiprocessors algorithm (up to few thousands of processors).
 In the example shown in Fig. \ref{fig:1-203} number of atoms is 192 millions, 1024 processors.
 Every processor simulates atom motion within its Voronoi subdomain and change a position of subdomain to achieve a good CPU load balance with neighbor subdomains.
 Here the map of subdomains is presented for the example shown in Fig. \ref{fig:1-203} for the same times:
  $\hat t=0.37$ (a), $\hat t=0.997$ (b), and $\hat t=1.62$ (c).
 We see how subdomains move and concentrate in the central zone where the massive jet absorbing mass of a shell is formed.
 (a) and (b) present the view from the top, while (c) shows the lateral view, comp. with Fig. \ref{fig:1-203}(c).
 All subdomains are shown, therefore the domains belonging to the same line of view overlap each other.
     }
      \label{fig:2-203}
\end{figure}

 As it is said in the caption to Fig. \ref{fig:2-203} our MD code is based on the Voronoi dynamic domain decomposition $(VD^3)$ approach.
 In this scheme the matter of the simulated sample is divided between the material particles (VS -- Voronoi subdomains).
 Each VS is linked to a single processor (or CPU core) to perform MD calculations.
 As the simulation progresses through dramatic changes within the material, the size and shape of each VS evolves in such a way
  as to keep good load balance.
 To do this the continuous current exchange between the neighboring processors is applied.
 In conditions of good load, there are approximately equal number of atoms in each VS.

 The division onto VS is accomplished by adaptive decomposition of the material into Voronoi subdomains (polyhedra).
 The full geometry of the decomposition, including the size and shape of individual Voronoi subdomains, is determined solely by positions of the VS centers
   through the standard construction algorithm of Voronoi tessellation.
 Then, the load balance in the course of simulation is controlled by the time evolution of the subdomain centers,
   which move according to a load-balancing algorithm that calculates the displacement of each center.
 The calculation of the displacement is based on differences in processor time per simulation step spent by the VS and its neighbors.

 Once displacements are calculated and the new dividing planes between subdomains are defined,
   the VS exchange atoms according to the new decomposition geometry.
 Such local load balancing results in good global balancing achieved within several time steps of the simulation.
 Our work demonstrates that the auto-balancing Voronoi decomposition is both a very efficient and fast algorithm.

 The difference between static and dynamic $VD^3$ decompositions can be nicely illustrated by analogy with the Euler (static decomposition)
  and Lagrangian $VD^3$ representations of fluid dynamics.
 The latter method, which follows the evolution of fluid particles,
   is obviously more effective for studies of inhomogeneous and rapidly evolving flows of matter.

 Example of the combined two-temperature hydrodynamics and MD simulations is shown in Figures \ref{fig:1-203} and \ref{fig:2-203}.
 Parameters of the MD part are: almost $0.2\cdot 10^9$ atoms, 1024 processors, the run covers 4 ns of evolution of a film,
  duration of the temporal step on computer is 1.2 s, $10^6$ temporal steps were done.
 This takes 2 weeks of CPU time;
   17000 kWh of electric energy were consumed by computer during the run.
 260 microW/atom is the "payment" for running of this simulation.

\begin{figure}[h]    
\includegraphics[width=1\textwidth]{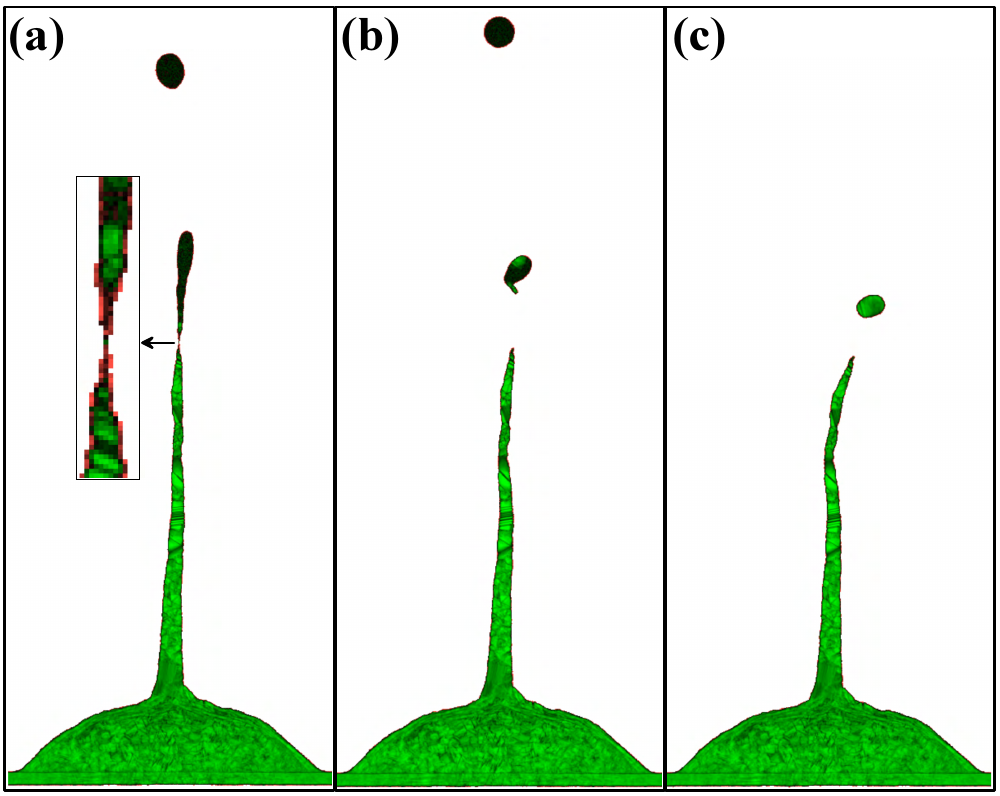}
\caption{ Final stages of formation of cupola and jet.
 (a) $\hat t = 5.25,$  (b) $\hat t = 5.73,$  (a) $\hat t = 8.$ See text for explanations.
      }
      \label{fig:3-225}
\end{figure}

 The simulation run presented in Figs. \ref{fig:1-203} and \ref{fig:2-203} corresponds to the case with suppressed heat conduction.
 It is very important to include solidification to describe final morphology of the solitary microstructure created by the tight laser impact.
 To do this the electron thermal transfer is inserted in the MD code.
 Calculations of heat transport are performed by the Monte-Carlo (MC) subroutine in the MD code.
 The MC subroutine and its introduction in the MD code (MD+MC) are described in \cite{Inogamov:2016b}.
 In the next Section we apply the MD+MC code to study development of the solid cupola and jet and to consider dynamics of decay of a jet.

 \section{Rupture of a jet in liquid and solid states}
 \label{sec5:jet.rupture}

 As was said, Figures \ref{fig:1-203} and \ref{fig:2-203} belong to the situation with negligible heat conduction.
 Therefore the flying film and jet for long time remain in a liquid state; radiative cooling is slow.
 Opposite case is presented in Fig. \ref{fig:3-225}.
 It is simulated by the MD+MC code.
 Here cooling is fast enough to stop a cupola in its flight.
 Solidification qualitatively changes the situation relative to the molten case.
 In Fig. \ref{fig:3-225} green colors correspond to recrystallized gold.
 This is rather late stage of evolution.
 Only the upper droplet in the top of the frames (a) and (b) in Fig. \ref{fig:3-225} is liquid.

 In this Section we consider decay of the solidified jet.
 Therefore the late instants $\hat t = 5.25\div 8.00$ of evolution are shown (run 225).
 At the early previous stages the cupola inflates, after that surface tension begins to decelerate inflation.
 The jet begins to grow after the stage when the cupola stops and turns back
        (thus the inflation is changed to deflation), see Section \ref{sec3:Capillary} above.
 A very long jet is developed during deflation.
 The Plateau-Rayleigh instability of a liquid column produces an approximately linear vertical chain of droplets.
 Velocities and temperatures of the successive droplets decreases -- the first one is the most fast and hot.
 Generation of droplets continues up to the stage and place on the column of the jet
   where the solidification zone moves up along the column.
 Thus crystallization begins to interplay with dynamics of rupture.


 The last nanoparticle separated from the jet (now the continuous rest of the jet transforms to the crystallized spike)
   separates in the semi-liquid -- semi-solid mixed state.
 Thus we cannot call it a liquid droplet.
 The upper part of the nanoparticle is liquid, see Fig. \ref{fig:3-225}(a),
   while the bottom part (which is near the neck of rupture) is a freshly crystallized solid.
 As was mentioned in papers \cite{Inogamov:2016a,Inogamov:2016b,Inogamov:2014c,Wu:2016a}, the recrystallization process after femtosecond pulse
   proceeds in strongly overcooled conditions: few hundreds Kelvins below melting temperature. 
 Later in time (after separation) the semi-solid nanoparticle demonstrates interesting thermal and mechanical behavior.

\clearpage
\begin{figure}[p]
\includegraphics[angle=90,origin=c,width=0.85\textwidth]{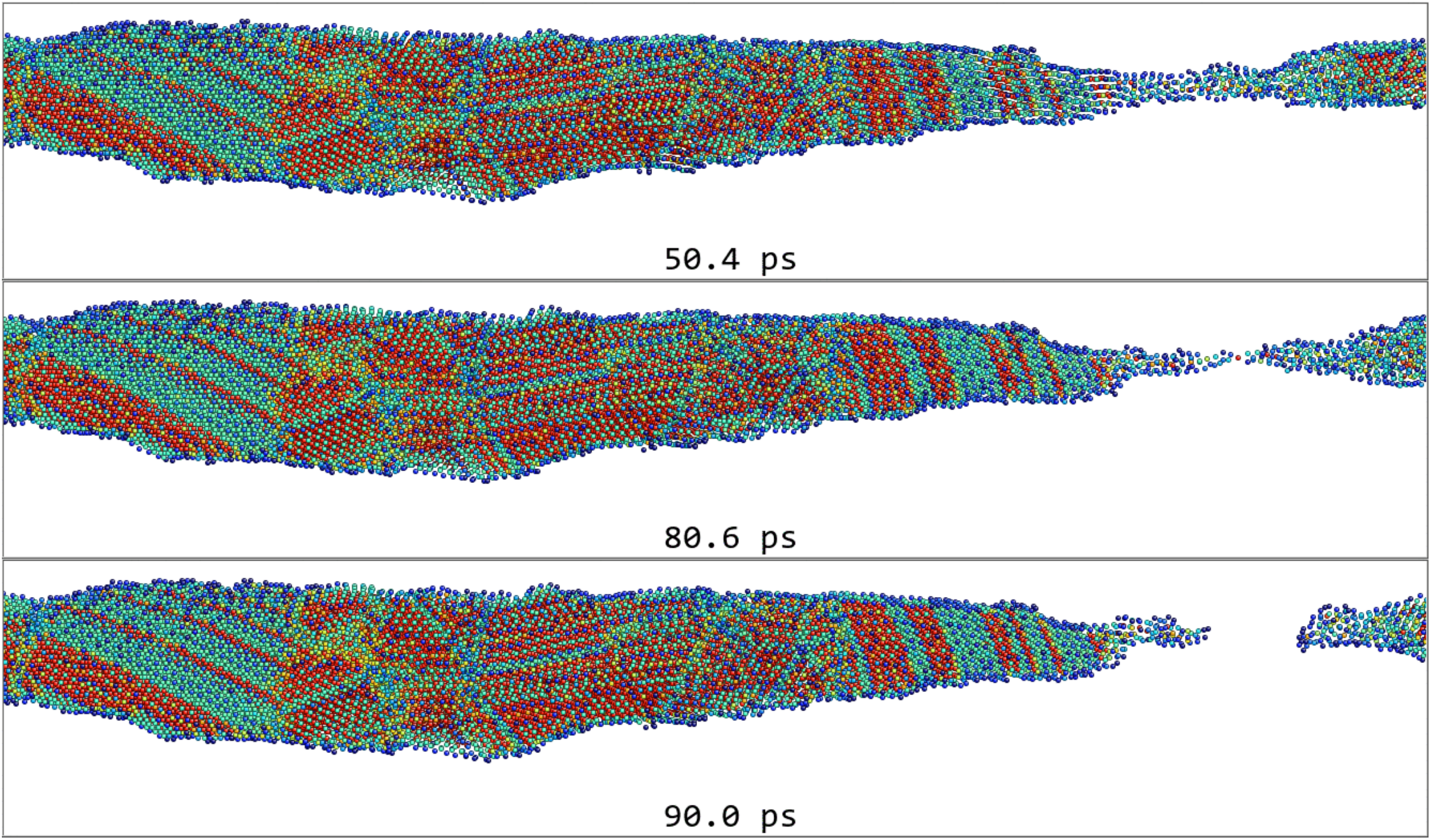}
\caption{ Atomic positions just before and just after the rupture of the solid part of a jet.
 The place of the rupture is marked by the arrow in Fig. \ref{fig:3-225}(a).
 Shown here evolution near the instant of rupture is much shorter than in Fig. \ref{fig:3-225}.
 We change the vertical position of the jet in Fig. \ref{fig:3-225} to the horizontal position here to save place.
 See the text.
     }
      \label{fig:4-225}
\end{figure}

 The thermal history is: the solid part of the nanoparticle is gradually covered by liquid:
  see in Fig. \ref{fig:3-225}b how the sharp solid tail (which was just connected to the neck) turns in space
   and is pulled by 
     surface tension into the rounded liquid part of the particle.  
 After that the whole particle solidifies because, as was said, it is in a strongly overcooled state.

 Let's describe mechanics.
 There is a continuous distribution of vertical velocity along a liquid part of a jet.
 In a solid part these velocities are significantly smaller.
 Vertical velocity $u$ is directed up (see Figs. \ref{fig:1-203}, \ref{fig:3-225})
  and its instantaneous values increases with the instantaneous heights of the liquid Lagrangian particles
   which compose the jet.
 Every liquid Lagrangian particle is approximately in the inertial flight thus keeping approximately its velocity $u.$
 Only the first (or the head) droplet slowly decreases its velocity $u$ due to action of the surface tension in the neck
  \cite{Inogamov:2015a} where the head droplet transits into cylindrical jet.

 Liquid near the freezing zone is stretched strongly - here $\partial u/\partial z$ is larger than in a liquid jet above the freezing zone.
 This local stretching is unstable - if some narrowing appears then it will become more narrow because of enhancement of stress at a smaller cross section.
 Thus the process of narrowing of a neck begins.
 Namely this process results in a rupture.
 Simultaneously the freezing zone moves up and passes the narrowing {\it before} the rupture.
 Therefore slightly later (than the passing of freezing through the neck) the rupture takes place in a {\it solid} state.

 Fig. \ref{fig:4-225} presents crystallographic structure of the solidified jet in vicinity of the neck;
   see also inset in Fig. \ref{fig:3-225}a where the neck is enlarged.
 As in Fig. \ref{fig:3-225} the picture shows distribution of the central-symmetry parameter $s$
   which is defined by positions of the neighboring atoms relative to the chosen one.
 But in Fig. \ref{fig:3-225} the parameter $s$ is averaged over the pixels
   while in Fig. \ref{fig:4-225} (using the package AtomEye) every atom is colored according to the value of $s.$
 We see formation of the crystalline strips oblique to the axis of the jet in Fig. \ref{fig:4-225}
   similar to those in Fig. 3c in \cite{Nakata:2013}.
 Below the tip the 5-fold symmetry polyicosahedral structures are formed.
 They are the same as in the jet in \cite{Wu:2016a}.
 But the jet in \cite{Wu:2016a} is formed after decay of membranes \cite{Inogamov:2014c}
  forming foam with small liquid content.
 While here the jet appears after the pointed laser impact.
 Mechanical properties of foam (value of negative pressure as a function of liquid content) were considered in
  \cite{Mayer:2016,Zhakhovskii:2008a}.
 The paper \cite{Zhakhovskii:2008a} was the first
  where an explanation why the surface nanostructures appears was given.
 Fig. \ref{fig:4-225} demonstrates break-off of the narrowing.
 This process produces a sharp crystalline tip.

\end{document}